\documentclass[twocolumn,aps,prb,showpacs]{revtex4}
\usepackage{epsfig,graphicx,times}

\usepackage{amssymb}
\usepackage{amsmath}
\usepackage{graphicx}
\usepackage{bm}

\begin{document}
\draft

\title{Photon-Induced Magnetization Changes in Single-Molecule Magnets}

\author{ M. Bal$^1$, Jonathan R. Friedman$^{1,*}$, E. M. Rumberger$^2$, S. Shah$^2$, D. N. Hendrickson$^2$, N. Avraham$^3$, Y. Myasoedov$^3$, H. Shtrikman$^3$ and E. Zeldov$^3$}
\affiliation{\mbox{$^1$ Department of Physics, Amherst College, Amherst, Massachusetts 01002-5000} \\
\mbox{$^2$Department of Chemistry and Biochemistry, University of California at San Diego, La Jolla, California 92093}
\mbox{$^3$Department of Condensed Matter Physics, The Weizmann Institute of Science, Rehovot 76100, Israel}}

\date{\today}

\begin{abstract}

Microwave radiation applied to single-molecule magnets can induce large magnetization changes when the radiation is resonant with transitions between spin levels.  These changes are interpreted as due to resonant heating of the sample by the microwaves.  Pulsed-radiation studies show that the magnetization continues to decrease after the radiation has been turned off with a rate that is consistent with the spin's characteristic relaxation rate.  The measured rate increases with pulse duration and microwave power, indicating that greater absorbed radiation energy results in a higher sample temperature.  We also performed numerical simulations that qualitatively reproduce many of the experimental results.  Our results indicate that experiments aimed at measuring the magnetization dynamics between two levels resonant with the radiation must be done much faster than the $\geq$20- $\mu$s time scales probed in these experiments.

\end{abstract}

\maketitle

Single-molecule magnets (SMMs) have been intensively studied for more than a decade.  There are multiple motivations for studying these systems.  SMMs are some of the smallest bistable magnetic units, exhibiting hysteresis below a blocking of a few Kelvin.  With each molecule in a crystal nominally chemically identical, they are better characterizable than almost all other systems of nanomagnets.  From a practical point of view, SMMs represent the ultimate in magnetic storage density.  At the same time, SMMs exhibit unique quantum signatures, such as tunneling between spin-orientation states (i.e.~m levels)\cite{33, 497, 81} and interference between tunneling paths.\cite{162}

In the last few years, several groups have been studying the interaction of SMMs with radiation in an attempt to control the magnetization dynamics\cite{283, 343, 339, 337, 562, 505, 560, 561}.  Beyond electron-spin resonance (ESR) on SMMs,\cite{88, 417} these experiments measure changes in the sample magnetization in the presence of continuous and pulsed microwave radiation.  Ideally, the use of radiation to manipulate the dynamics has the advantage that population is moved between only two levels - those resonant with the applied radiation - while other control parameters, such as temperature, affect the population dynamics of multiple levels.  Motivation for these experiments comes from the possibility that these systems might be used for quantum computing, as has been explored theoretically.\cite{298}  For that purpose, it is essential to measure the lifetime $T_1$ of excited states, as well as their decoherence times $T_2$, both of which can in principle be measured with time-domain techniques.  Measuring these quantities will allow for a deeper understanding of how the molecular magnets interact with their environment.  For example, spin-phonon interactions most likely limit the lifetime of the excited states.  However, these interactions are still not fully understood.  Calculations of the relaxation rate of the molecular magnets using standard spin-phonon interactions yield results that are orders of magnitude slower than measured.\cite{221}  On the other hand, including interactions that allow a phonon to induce $\Delta m$ = 2 changes in spin give results that are in reasonable agreement with the data;\cite{113, 168} however, the existence and amplitude of such interactions have been debated and other mechanisms have been proposed.\cite{205, 567}  Thus, a direct measurement of the excited-state lifetimes may help shed light on the actual spin-phonon relaxation mechanisms in the SMMs.  More generally, an understanding of decoherence in the SMMs will help address the question of how objects lose their quantum properties in the macroscopic limit.

In this paper, we present the results of studies of the magnetization dynamics of SMMs in the presence of resonant microwave radiation. Our results indicate that large magnetization changes can be induced in SMMs by the application of resonant microwaves.  However, much of these changes are due to resonant sample heating by the microwaves on the time scale of the spin's characteristic relaxation time.

\begin{figure}[htb]
\centering
\includegraphics[width=80mm]{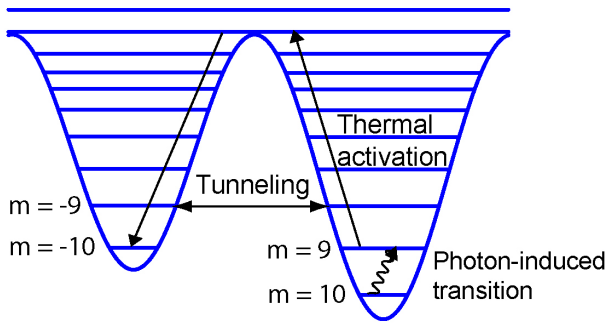} 
\caption{Double-well potential for a typical SMM.  The right well corresponds to the spin pointing ``up" and the left corresponds to it pointing ``down".  A magnetic field tilts the potential, making one well higher than the other.  The energy levels shown are for the Fe$_8$ SMM but the level distribution is qualitatively similar for other SMMs.  Mechanisms by which magnetization can be changed are illustrated by the arrows.  Tunneling is generally only significant when levels in opposite wells are aligned, as in the figure.
\label{doublewell} }
\end{figure}

The spin dynamics of SMMs can generally be described by the effective Hamiltonian

\begin{equation}
{\cal H} =  - DS_z^2  - BS_z^4 - g\mu _B S_z  H_z +{\cal H'}
\label{Ham}
\end{equation}

\noindent where ${\cal H'}$ contains terms that do not commute with $S_z$ and therefore produce tunneling, and $D$ and $B$ are anisotropy constants.  The first term impels the spin to align with or opposite the (easy) z axis, giving rise to a double-well potential, illustrated in Fig.~\ref{doublewell}.  The energy levels are then approximately the eigenstates of $S_z$, with magnetic quantum numbers $m$, as indicated in the figure.  When a magnetic field $H_z$ is applied along the easy axis, it tilts the potential, lowering the energy of, say, the right "up" well while raising the energy of the opposite well.  Resonant tunneling occurs when the field causes levels in opposite wells to align.

\begin{figure}[htb]
\centering
\includegraphics[width=79mm]{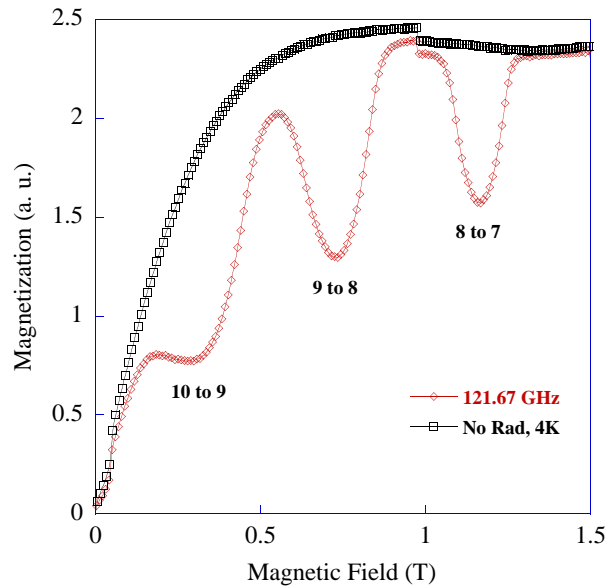} 
\caption{Magnetization as a function of field for the Fe$_8$ SMM both in the presence of 121.67-GHz cw microwave radiation and with no radiation.  The radiation heats the sample, especially when it is resonant with a transition between spin energy levels, producing the observed dips.  The cavity Q is $\sim$2000 and the  power incident on the cavity (after accounting for losses in the waveguide and other components) is $\sim$3 mW.
\label{dips} }
\end{figure}

Many of our studies have focused on the so-called Fe$_8$ molecular magnet (Fe$_8$O$_2$(OH)$_{12}$(tacn)$_6$) for which $S$ = 10, $D=$ 0.292 K and $B$ is negligible.  When continuous wave microwave radiation is applied dips occur in the magnitude of the equilibrium magnetization whenever the field brings a pair of neighboring levels into resonance with the radiation.  These results represent a variation on traditional ESR spectroscopy.  In our first experiments, using low-amplitude radiation, we found very small changes (less than 1\%) in the magnetization for the 10-to-9 transition.\cite{337}  In the current study, we used a high-power backward-wave oscillator and a resonant cavity consisting of a two-inch piece of WR-6 waveguide terminated on both ends with gold or copper foil.  A small coupling hole at one end allowed radiation to couple into the cavity from gold-plated stainless steel WR-10 waveguide that piped the microwaves into the cryostat.  Depending on the positioning of the cavity caps and the mode excited, the Q could be as high 2500.  The sample was mounted on a thinned side wall of the cavity, opposite a Hall sensor located outside the cavity, and oriented with its a axis parallel to the field direction, resulting the easy axis being tilted by $\sim$16$^\circ$ relative to the field.  Samples were on the order of 100 $\mu$m in the longest dimension, much smaller than the wavelengths used.  Samples were mounted using silicone-based vacuum grease, which has a low thermal conductivity.  (The sample was cooled indirectly by a few Torr of helium exchange gas filling the sample chamber.)  With this new experimental configuration, we observed very large photon-induced changes in magnetization;\cite{562} an example is shown in Fig.~\ref{dips}, where several transitions are observed, as indicated.  There are two noteworthy features of this data.  First, the dips are much larger than one would first expect:  in the limit in which the 10-to-9 transition is saturated, at most 50\% of the initial ground-state ($m$ = 10) population is transferred into the excited ($m$ = 9) state, producing a maximum 5\% decrease in the magnetization, while the observed dips are much larger.  Second, we observe dips that correspond to transitions between excited states.  The sample temperature when the microwaves are not on resonance with a transition is $\sim$4K, smaller than the energies of the excited states.  Thus, the populations of the excited states should be small and all of the transitions between them should have a significantly smaller amplitude than the 10-to-9 transition.  Instead, all of the dips have similar amplitudes and all are larger than would be expected even if the transitions were saturated.

We surmise, therefore, that magnetization changes must involve population being moved to levels other than the two involved in the radiative transition.  In fact, it is likely that population is moving to the opposite metastable well.  The simplest explanation for our observations is that the radiation heats the sample significantly when it is on resonance with a transition between levels.  The dips in magnetization then represent the simple fact that the equilibrium magnetization decreases at higher temperatures.  We have seen direct evidence for such heating using a thermometer to monitor the cavity temperature and observed an increase in temperature whenever one of the magnetization dips occurs.\cite{562}  The fact that the magnetization dips occur even for transitions between excited states we believe is due to a bootstrapping process in which radiative transitions heat the sample, increasing the excited-state populations and, in turn, allowing for more photons to be absorbed.

\begin{figure}[htb]
\centering
\includegraphics[width=80mm]{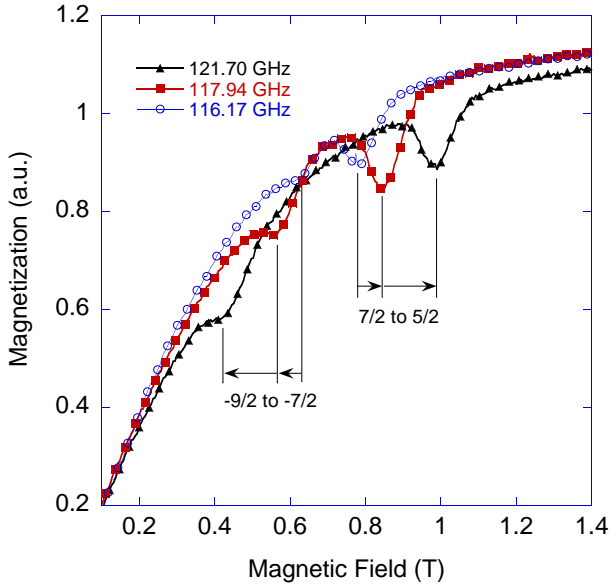} 
\caption{Equilibrium magnetization as a function of field when cw microwave radiation is applied to the Mn$_4$ SMM.  Data for three different frequencies is shown, as indicated.  The nominal system temperature is $\sim$ 2.4 -- 2.5 K.  At each frequency, two dips are observed, corresponding to the transitions indicated.
\label{Mn4} }
\end{figure}

Further evidence that the observed magnetization dips are due to radiative heating is provided by a study of another SMM, Mn$_4$O$_3$Cl(O$_2$CCH$_3$)$_3$(dbm)$_3$, or simply Mn$_4$ for short.  This molecule, with a spin of 9/2 and $D=$ 0.76 K, has a smaller barrier than Fe$_8$.  For these samples, the easy axis was closely aligned with the magnetic-field direction.  Equilibrium magnetization curves taken in the presence of microwaves at three different frequencies are shown in Fig.~\ref{Mn4}.  Two sets of dips occur.  One dip moves to higher fields with increasing frequency, consistent with a transition between levels in the right well, where levels get further apart with increasing field.  We identify this dip with the 7/2-to-5/2 transition.  The other dip moves to lower fields with increasing frequency, indicating that it belongs to a transition in the left well, which we identify as a transition between m = -9/2 and m = -7/2.  Since this latter transition occurs in the metastable well, if radiation were only transferring population between the two states of the transition, the equilibrium magnetization should increase in the presence of radiation.  The fact that we instead observe a decrease in magnetization for this transition indicates that the primary effect of the radiation is heating of the sample.

\begin{figure}[htb]
\centering
\includegraphics[width=80mm]{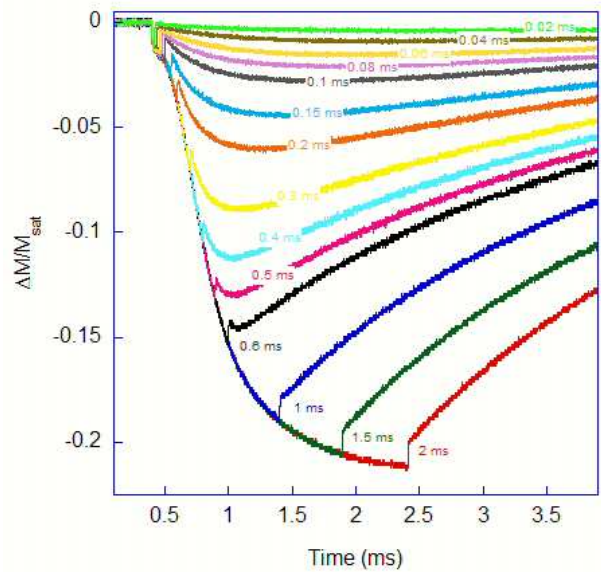} 
\caption{Magnetization change (normalized to the saturation magnetization) as a function of time when pulses of 119.77 GHz radiation are applied to the Fe$_8$ SMM.  Data for various pulse durations, as indicated, are shown.  The abrupt jumps when the radiation is turned on or off are due to an instrumental artifact.  The nominal system temperature is 1.8 K and the applied magnetic field is 0.17 T, at which the 10-to-9 transition is resonant with the radiation.  The radiation power incident on the cavity was 6.2 mW and the cavity Q was several hundred.
\label{pulselength} }
\end{figure}

To better understand how this heating takes place, we performed a study of the magnetization dynamics when pulses of radiation were applied to the Fe$_8$ SMM.  Fig.~\ref{pulselength} shows the change in magnetization (normalized by the sample's saturation magnetization) at (nominally) 1.8 K and 0.17 T as a function of time during and after 119.77 GHz (corresponding to the 10-to-9 transition) radiation pulses of various durations, as indicated.  The magnetization decreases during the radiation pulse, following a ``backward-S"-shaped curve, seen for the longest pulses. (The data exhibit abrupt jumps whenever the radiation is turned on or off, which are due to an instrumental artifact.)  In addition, except for the longest pulses, the magnetization continues to decrease after the radiation has been turned off and eventually reaches a minimum. These effects are not observed (or at most are very weak) when the radiation is not resonant with a transition between spin states.\cite{562}  We interpret the results as being due to an indirect heating of the spin system by the radiation.  As described previously,\cite{562} after photons are absorbed by the spin transition, the spins relax through the emission of phonons. These phonons then rapidly thermalize, heating the lattice to a higher temperature.  The spin system must then come to thermal equilibrium with the lattice, a process that takes the spin's characteristic relaxation time for thermal activation and/or thermally assisted tunneling.  At the temperatures of our experiments, the relaxation time for SMMs is known to follow an Arrhenius law

\begin{equation}
\tau=\tau_0 e^{-U/k_BT},
\label{arr}
\end{equation}

\noindent where $U$ is the effective activation barrier and $\tau_0 \sim 10^{-7}$ s for many SMMs including Fe$_8$.  The backward-S shape can be qualitatively interpreted as follows.  At short times ($\lesssim$ 0.1 ms), the magnetization does not change very much because (a) the amount of energy absorbed by the crystal is small and (b) the spins have not yet had enough time to come to thermal equilibrium with the new lattice temperature.  At somewhat longer times ($\sim$0.1  -- $\sim$1 ms), the magnetization decreases rapidly as the spins try to reach thermal equilibrium with the higher-temperature lattice.  The spins can relax increasingly fast at progressively longer times as the temperature of the lattice continues to increase during the pulse.  At the longest times ($\gtrsim$ 1 ms), the magnetization begins to saturate as the entire system (spins and lattice) reach a well-defined steady-state temperature and the rate of radiation energy absorption is balanced by the rate heat to be carried away by the cryostat.  For the longest pulse durations, the magnetization does not decrease after the radiation has been turned off, consistent with the interpretation that the spins are close to equilibrium with the lattice at these time scales.

\begin{figure}[htb]
\centering
\includegraphics[width=80mm]{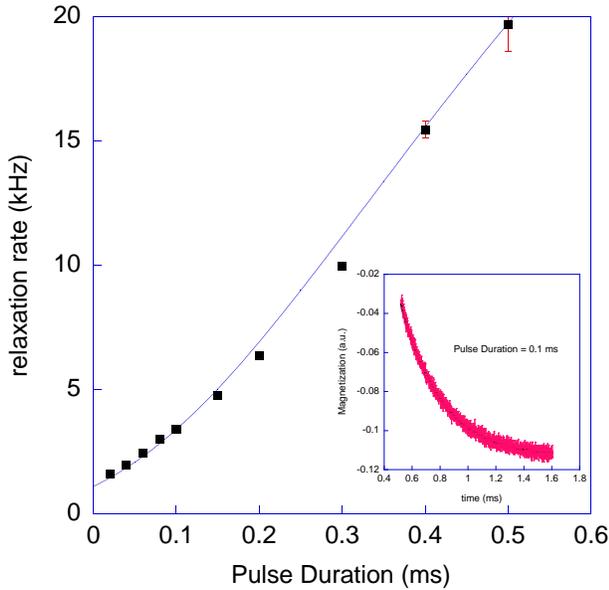} 
\caption{Relaxation rate as a function of pulse duration for the Fe$_8$ SMM.  The rates are extracted by fitting the post-pulse decays in Fig.~\ref{pulselength} to an exponential function, an example of which is shown in the inset.  The curve in the main figure is the result of numerical calculations, as described in the text.
\label{ratevlength} }
\end{figure}

A more quantitative analysis of the data can be carried out using the decrease in magnetization that occurs after the radiation has been turned off.  The inset of Fig.~\ref{ratevlength} shows a typical such decrease, as well as a fit to an exponential decay.  The fact that such a fit works well is notable, indicating that during this period, the lattice temperature is reasonably constant so that the spins have a single definable relaxation rate.  This is confirmed by the fact that the time for the magnetization to return to equilibrium after the pulse is much longer than the timescale of the decay.  In previous work, we used such fits to extract relaxation rates ($1/\tau$) from the post-pulse decays and found that the rate is maximum at fields for which the absorption for the 10-to-9 transition is strongest or for which resonant tunneling is active, confirming that the rate corresponds to the spin's relaxation across the barrier.\cite{562}  Here we apply a similar analysis to the data in Fig.~\ref{pulselength} to see how the rate depends on pulse duration; the results are shown in the main part of Fig.~\ref{ratevlength}.  There is a clear trend to increasing relaxation rate for longer pulse lengths.  This is consistent with more heating with longer pulses.  Extrapolating the data in Fig.~\ref{ratevlength} to zero pulse duration yields a relaxation rate of $\sim$1 kHz.  The value is roughly an order of magnitude faster than other published measurements of the relaxation rate in this system,\cite{193, 91} although these do not report values for our precise experimental conditions (1.8 K and 0.17 T).

\begin{figure}[htb]
\centering
\includegraphics[width=80mm]{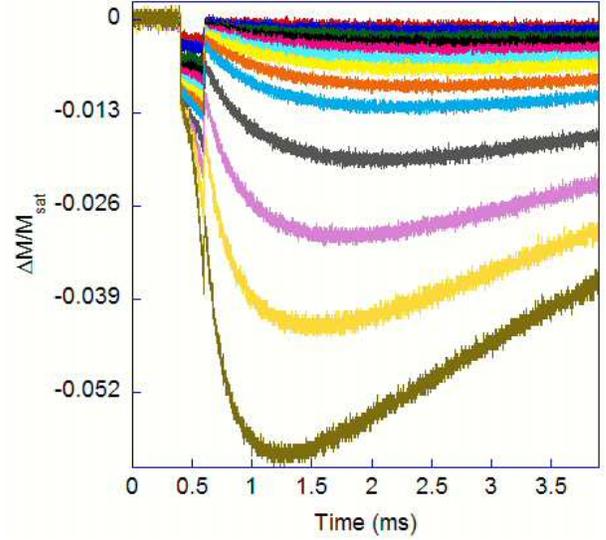} 
\caption{Magnetization change as a function of time when 0.2-ms pulses of 119.77-GHz radiation are applied to the Fe$_8$ SMM.  From top to bottom, the curves correspond to power levels of 0.21 mW, 0.34 mW, 0.59 mW, 0.76 mW, 1.00 mW, 1.24 mW, 1.48 mW, 1.86 mW, 2.24 mW, 3.00 mW, 3.96 mW, 4.96 mW and 6.20 mW.  The cryostat temperature is 1.8 K, magnetic field 0.17 T and cavity Q several hundred.
\label{power} }
\end{figure}

\begin{figure}[htb]
\centering
\includegraphics[width=80mm]{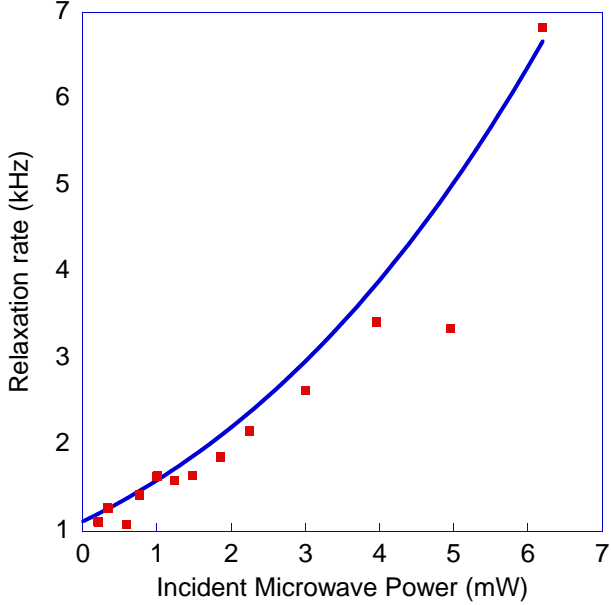} 
\caption{Relaxation rate as a function of microwave power extracted from the post-pulse decays in Fig.~\ref{power}.  The solid curve is the result of numerical calculations, as described in the text.
\label{ratevpower} }
\end{figure}

Figure \ref{power} shows 0.2-ms-pulsed-radiation data for several values of radiation power.  The behavior is qualitatively similar to that in Fig.~\ref{pulselength}:  the magnetization decreases during the pulse, then continues to decrease after the pulse, reaching a minimum, before slowly returning to equilibrium.  At higher powers the magnetization changes are larger, but the behavior does not change qualitatively.  We again analyze the decay after the pulse by fitting to an exponential.  The relaxation rate extracted this way is shown in Fig.~\ref{ratevpower}.  The rate increases with power and has a zero-power intercept of just above 1 kHz, similar to what is found from the pulse-duration dependence.  The observed increase in relaxation rate is in qualitative agreement with our interpretation: the higher the radiation power, the higher the lattice temperature and the faster the spins relax.

It is interesting to note that Petukhov {\em et al}.,\cite{560} also working with Fe$_8$ do not see large decreases in the magnetization after the radiation is turned off.  The reason for the disparity been our results and theirs is unclear, but may have something to do with thermal anchoring of the sample.

We simulated the non-equilibrium dynamics of our system by tracking the population of the ground state as it is driven out of equilibrium by the heating of the lattice and attempts to regain equilibrium through thermal relaxation.  We first calculate the photon-induced transition rate for the 10-to-9 transition.  This can be done using the standard expression\cite{86}

\begin{equation}
w = \frac{{\pi H_1^2 }}{{2\hbar ^2 }}\left| \mu  \right|^2 F(\omega ),
\end{equation}

\noindent where $\left| \mu  \right|^2  = (g \mu_B )^2 \left| {\left\langle 10 \right|(S_ +   + S_ -)/2  \left| 9 \right\rangle } \right|^2 $ is the square of the magnetic dipole matrix element, $H_1$ is the amplitude of the radiation field and $F(\omega)$ is the lineshape function, which has a value of $\pi/T_2$ for the peak of a Lorentzian.  We estimate $H_1$ inside the cavity to be $\sim$0.06 Oe for an incident power of 6.2 mW and cavity Q$\sim$370.  $T_2$ is $\sim1.7 \times 10^{-10}$ s.\cite{337}  (While this small value likely represents inhomogeneous broadening, the nature of the broadening is immaterial to the present calculation as long as the transition is not driven near saturation.)  With these numbers, we calculate a transition rate of $w$$\sim$400 s$^{-1}$.  Since this value is much smaller than any spin-phonon transition rate, we can conclude that photon absorption alone does not appreciably change the populations of the states.  So, we treat the levels within the right well as all being in thermal equilibrium with the lattice during the radiation pulse.  The magnetization dynamics are then driven by the radiative heating of the lattice.  As the system heats, the (normalized) population of the ground state $p_{10}$ relaxes towards thermal equilibrium through thermal activation over the barrier with a time constant $\tau$ given by Eq.~\ref{arr}:

\begin{equation}
dp_{10}/dt = -(p_{10}-p_{10}^{eq}(T))/\tau,
\label{popeq}
\end{equation}

\noindent where $p_{10}^{eq}(T)=e^{-E_{10}/k_BT}/Z$ is the equilibrium population of the ground state at temperature $T$ and $Z=\sum\limits_i {e^{-E_i/k_BT}}$ is the partition function for the spin system.  The energy eigenvalues $E_i$ are found by numerical diagonalization of the Hamiltonian.

The temperature of the system as a function of time can be found through energy considerations.  The rate of change of the sample's energy can be related to the energy of the photons absorbed:

\begin{align}
dW/dt &= w\hbar \omega N (p_{10}-p_9) - \kappa(T-T_{env}) \nonumber \\
&= w\hbar \omega N p_{10}(1-e^{-(E_9-E_{10})/k_BT}) - \kappa(T-T_{env}),
\label{heat}
\end{align}


\noindent where $N\sim 1.7 \times 10^{14}$ is the total number of spins in the sample.  The last term represents the heat flow out of the sample to the cryostat at temperature $T_{env}$ (1.8 K for our experimental situation); $\kappa$ is a phenomenological parameter characterizing the thermal conductance.  The energy gain by the sample can be related to its temperature change by

\begin{equation}
dW = m C dT \Rightarrow dW/dt = m C dT/dt,
\label{Cheat}
\end{equation}

\noindent where $m\sim$ 0.5 $\mu$g is the sample mass.  The specific heat $C$ has two contributions.  The lattice specific heat is\cite{280} $C_{lat} = 2.1 \times 10^{-5} T^3 J/gK^4$.  The magnetic specific heat (per molecule) is calculated from the usual definition:  $C_{mag} = k_B\frac{\partial}{\partial T}(T^2 ~\frac{\partial lnZ}{\partial T})$.  $C_{mag}$ is not strictly defined when the spins are out of equilibrium.  In reality, at low temperatures only states in the lower well are in thermal equilibrium with the lattice while states in the other well are inaccessible, effectively lowering the specific heat until the spin system comes into equilibrium.  An effective time-dependent specific heat has been considered theoretically.\cite{280, 123, 156, 138}  We chose not to include such a careful accounting of the magnetic specific heat because it both significantly increases the complexity of our calculations and results only in a factor of $\sim$2 decrease in the specific heat during the time when the spins are out of equilibrium.

\begin{figure}[htb]
\centering
\includegraphics[width=80mm]{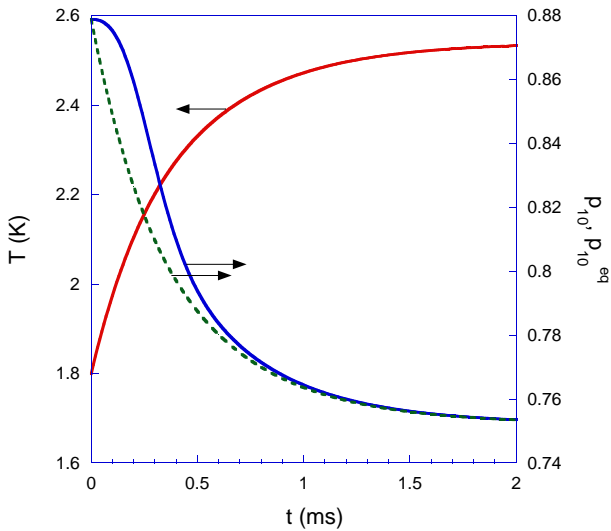} 
\caption{Results of numerical simulations for the Fe$_8$ SMM in the presence of radiation resonant with the 10-to-9 transition.  The temperature and ground-state population are shown as a function of time, with resonant radiation applied continuously starting at $t = 0$.  The dashed curve shows the equilibrium value for the ground state population $p_{10}^{eq}$.
\label{simresults1} }
\end{figure}

Equating Eqs.~\ref{heat} and \ref{Cheat}, results in an equation for $dT/dt$.  This equation and Eq.~\ref{popeq} can be simultaneously solved numerically to find $p_{10}$ and $T$ as a function of time.  From $T(t)$ we can calculate the relaxation rate ($1/\tau$) of the systems after a pulse of length $t$ using Eq.~\ref{arr}.  In doing these calculations, we used experimentally determined values of the necessary parameters (radiation power, cavity Q, sample mass, etc.).  We set the energy barrier $U$ to be 22.7 K, the calculated height of the classical barrier at a field of 0.17 T.  In order to obtain a relaxation rate in the absence of radiation of $\sim$1 kHz (the intercepts in Figs.~\ref{ratevlength} and \ref{ratevpower}) we set $\tau=3 \times 10^{-9}$ s, an order of magnitude smaller than the published value.\cite{91}  We chose $\kappa = 5 \mu$J/Ks, which results in good agreement between the measured dependence of the relaxation rate on pulse duration and the calculated values, as shown in Fig.~\ref{ratevlength}.  Given the freedom in these parameters, the agreement is reasonably satisfactory.  Without adjusting these parameters, we were also able to simulate the relaxation rate dependence on incident radiation power and achieve a good agreement with the data, as shown by the solid curve in Fig.~\ref{ratevpower}.  Having fixed the parameters, we calculated $p_{10}$ and $T$ as a function of $t$.  The results of these calculations are shown in Fig.~\ref{simresults1}.  These results show that the (lattice) temperature rises quickly and then saturates at $\sim$2.5 K.  More interestingly, $p_{10}$ shows qualitative agreement with the data in Fig.~\ref{pulselength}, exhibiting the characteristic backward-S shape and reaching saturation on a time scale of $\sim$1 ms.  The maximum relative change in $p_{10}$ is 12.5 \%, which is roughly half the maximum relative change in magnetization in Fig.~\ref{pulselength}, as expected since $\Delta M/M_{sat} \simeq \Delta (p_{10}-p_{-10}) \simeq 2 \Delta p_{10}$.  For reference, the figure also shows $p_{10}^{eq}$ as a dashed curve.  The difference between $p_{10}$ and $p_{10}^{eq}$ indicates that the spin system is out of equilibrium for $\sim$1 ms, again in fair agreement with with our observations.  A more quantitative comparison between the data and calculations would require inclusion of a proper treatment of the specific heat in the calculations as well as a better determination of some relevant experimental parameters, such as the relaxation rate in the absence of radiation.

In conclusion, our results indicate that the magnetization dynamics induced by microwaves on the time scales of our experiments is dominated by resonant sample heating.  The timescale for the spin system to respond to such heating corresponds to the spin's characteristic relaxation time.  In order to observe the pure radiation-induced dynamics between the two levels involved in the transition, experiments need to be done much faster than this relaxation time.  In fact, since phonons can only be emitted on the time scale of $T_1$, all heating effects could be avoided by working at time scales shorter than this.  The Hall sensors used in the work described herein have a response time of a few microseconds, which is unlikely to be fast enough.  An inductive pick-up loop is currently being used as a fast detector of magnetization changes. Preliminary results of that study are discussed elsewhere in this volume.

We thank Y. Suzuki for her work on this project and D. Krause and G. Gallo for their technical contributions.  Support for this work was provided by the National Science Foundation under grant number CCF-0218469, the Research Corporation, the Alfred P. Sloan Foundation, the Center of Excellence of the Israel Science Foundation, and the Amherst College Dean of Faculty.

\end{document}